\documentclass[aps,amsmath,prx,reprint,groupedaddress]{revtex4-1}
\usepackage{graphicx}
\usepackage{dcolumn}% Align table columns on decimal point
\usepackage{bm}% bold math
\usepackage{amsfonts}
\usepackage{array}
\usepackage[colorlinks,linkcolor=blue,anchorcolor=blue,citecolor=blue]{hyperref}

\begin{document}

\title{One-Time Shot-Noise Unit Calibration Method for Continuous-Variable \\ Quantum Key Distribution}
\author{Yichen Zhang$^{1}$}
\author{Yundi Huang$^1$}
\author{Ziyang Chen$^2$}
\author{Zhengyu Li$^{2}$}
\author{Song Yu$^{1}$}
\thanks{yusong@bupt.edu.cn}
\author{Hong Guo$^{2}$}
\thanks{hongguo@pku.edu.cn}

\affiliation{$^1$State Key Laboratory of Information Photonics and Optical Communications, Beijing University of Posts and Telecommunications, Beijing 100876, China }

\affiliation{$^2$State Key Laboratory of Advanced Optical Communication Systems and Networks, Department of Electronics, and Center for Quantum Information Technology, Peking University, Beijing 100871, China}

\date{\today}

\begin{abstract}
The shot-noise unit in continuous-variable quantum key distribution plays an important and fundamental role in experimental implementation as it is used as a normalization parameter that contribute to perform security analysis and distill the key information. However, the traditional calibration procedure and detector model can not cover all system noise in practical application, which will result in some loopholes and influence the practical security. What's more, the traditional procedure is also rather complicated and has difficulty in compatible with automatic operating system. In this paper we propose a calibration model based on the proposed trusted detector model, which could naturally close the loopholes in practical application. It can help identify the shot-noise unit in only one step, which can not only effectively simplify the evaluation process but also reduce the statistical fluctuation, while two steps are needed in traditional method. We prove its feasibility and derive the complete version of the corresponding entanglement-based model. Detailed security analysis against arbitrary collective attacks and numerous simulation results in both the asymptotic limit regime and the finite-size regime are provided. A proof-of-principle experiment has been implemented and the results indicate that the one-time-calibration model can be employed as a powerful substitution to calibrate the shot-noise unit. Our method paves the way for the deployment of continuous-variable quantum key distribution with real time calibration and automatic operation.
\end{abstract}

\maketitle

%   ============================================================
%==================   Section one: Introduction   ==================
%   ============================================================

\section{Introduction}

Quantum key distribution (QKD) is designed with the aim of realizing a physical-principle guaranteed secure key distribution between the two legitimate parties: Alice and Bob ~\cite{Gisin_RevModPhys_2002,Scarani_RevModPhys_2009,Pirandola_RevModPhys_2019}. Continuous variable (CV) QKD~\cite{Weedbrook_RevModPhys_2012,Diamanti_Entropy_2015} is developed little posterior to discrete variable QKD but becomes more appealing by virtue of its adaptability of implementing in existing commercial telecom systems. CV-QKD protocols using coherent states~\cite{Grosshans_PhysRevLett_2002, Weedbrook_PhysRevLett_2004} is considerably simple to implement and reverse reconciliation of CV-QKD protocols can break through the ``3dB" limitation in reconciliation process~\cite{Grosshans_Nature_2003}; thus they are generally applied in most of the experiment demonstrations. Moreover, finite-size effect has also been extensively studied as its impact commonly influences the practical CV-QKD system performance~\cite{Leverrier_PhysRevLett_2013,Leverrier_PhysRevLett_2015,Leverrier_PhysRevLett_2017}. The maximum achievable secret key rate of QKD has also been investigate as the Pirandola-Laurenza-Ottaviani-Banchi (PLOB) bound\cite{Pirandola_NatCom_2017, Pirandola_QST_2018}, i.e., the fundamental limit of repeaterless quantum communications.

The use of a Gaussian-modulation coherent state with a CV-QKD protocol can achieve a relatively high secret key rate, and its security has been proved against arbitrary attacks in both asymptotic regime and finite-size regime~\cite{Leverrier_PhysRevLett_2013, Leverrier_PhysRevLett_2015, Leverrier_PhysRevLett_2017, Grosshans_PhysRevLett_2005,Navascues_PhysRevLett_2005,Pirandola_PhysRevLett_2009,Renner_PhysRevLett_2009}; thus is getting more popular in recent years. Experimental demonstrations based on laboratory conditions have been conducted to prove its feasibility, and field tests based on real-life environmental conditions are carried out subsequently for the future practical applications~\cite{Lance_PhysRevLett_2005,Lodewyck_PhysRevA_2007,Qi_PhysRevA_2007,Khan_PhysRevA_2013, Huang_SR_2016}, CV-QKD set-up based on the integrated silicon photonic chip platform have also been developed~\cite{Zhang_NP_2019}. Recently, the experimental results of long distance CV-QKD over $202.81$ {\rm km} of ultralow-loss optical fiber has been reported~\cite{Zhang_Arxiv_2020}.What is more, this CV-QKD system has also shown the advantages in metropolitan field tests over $50$ {\rm km} commercial fibers~\cite{Zhang_QST_2019}. Recently, great efforts made in the proving of the security of CV-QKD with discrete modulation and several progress has achieved~\cite{Arxiv_Zhang_2018,Ghorai_PhysRevX_2019,Lin_PhysRevX_2019}.

The shot-noise unit (SNU) can be vital for a CV-QKD system, as the calibrated SNU is treated as a normalization parameter in quantizing the quadrature measurement results, and this eventually contributes to the estimate of the secret key rate.Previous experiment demonstrations usually applied the two-time-evaluation (TTE) procedure~\cite{Lodewyck_PhysRevA_2007, Khan_PhysRevA_2013}, which requires first measuring the electronic noise of a practical homodyne detector and then measuring the output of the homodyne detector with the local oscillator (LO) path included. In this way, the SNU is calibrated by using the results of the second measurement minus the electronic noise obtained from the first measurement, which is obviously a rather complicated procedure. Since the SNU is not measured directly, this will certainly introduce more inaccuracy.

However, such a calibration scheme can open up security loopholes that an eavesdropper Eve can utilize to procure information about the key~\cite{Ferenczi_OSA_2007, Jouguet_PhysRevA_2013}. Eve can take actions to change the SNU during the key distribution procedure, and then the SNU used to normalize the measured quadratures will not be the same as the real SNU; in this way, Alice and Bob are liable to underestimate the channel excess noise, which then threatens the security of the CV-QKD system. Also, imperfections in the homodyne detection can also affect the evaluation of the SNU. Adopting SNU monitoring is a commonly used countermeasure against such attacks, while the local-local-oscillator (LLO) scheme, where the local oscillator is generated by Bob using an independent laser, can effectively resist attacks against the calibration of the SNU~\cite{Qi_PhysRevX_2015, Soh_PhysRevX_2015}.
%However, the LLO scheme has not been widely enforced yet and the LO monitor will inevitably increase the system complexity.

In this paper, we propose a one-time calibration model that uses two beam splitters to imitate the imperfections of a homodyne detector in a corresponding entanglementbased (EB) model. The new calibration model can simplify the calibration procedure, as it does not require measuring the electronic noise for each use of a practical system. We simply measure the output of the homodyne detector with the LO path connected, and take that measurement result as the new SNU. Also, the statistical fluctuations in the SNU introduced by the calibration procedure can be reduced by applying the one-time calibration model. Detailed analyses of the secret-key-rate calculations and the performance of the proposed model are fully provided. Finally, we consider the proposed model in the finite-size regime to test its performance in a more practical environment. We remark that the one-time calibration model can, basically, approach the system performance of the original two-time calibration model in both the asymptotic regime and the finite-size regime. A demonstration experiment is conducted in order to indicate its possibility, and the results prove that the one-time calibration model provides us with a better alternative for performing calibration procedures.

The rest of the paper is organized as follows: In Sec.~\ref{sec:2}, we review the definition of SNU then provide introduction of the conventional method applied in estimating SNU and the limitations exist under such method.
In Sec.~\ref{sec:3} we propose the complete one-time-calibration procedure where we start from the practical output of the homodyne detector.
In Sec.~\ref{sec:4}, we mainly focus on the performance of the one-time-calibration model, where we present the key rate calculation of the model in detail, and then analyze the model behaviour under finite-size regime. We also provide multiple simulation results and give meticulous discussions in this section. We provide experiment details in Sec.~\ref{sec:5} and the conclusions are drawn in Sec.~\ref{sec:6}.

\section{\label{sec:2} Calibration of shot-noise unit}
In this section we first review the trusted noise and trusted modelling of the homodyne and heterodyne detector, then describe the conventional approach of calibrating the SNU. Lastly, we point out the limitations in the existing SNU calibration method.

\subsection{Trusted noise and trusted detector modelling}
We start by reviewing the trusted noise modelling of the homodyne and heterodyne detectors.
Usually a homodyne detector or a heterodyne detector has two main imperfections: a finite detection efficiency $\eta$, and electronic noise ${\varepsilon _{ele}}$~\cite{Lodewyck_PhysRevA_2007}. In untrusted modeling, the imperfections of practical homodyne and heterodyne detectors contribute to the channel loss or the channel excess noise, which can be controlled by the eavesdropper Eve. Another way of modeling a homodyne or heterodyne detector is trusted modeling, based on the assumption that the apparatus of Bob¡¯s setup is not accessible to Eve~\cite{Fossier_JPB_2009}. This way we can consider the imperfections of the homodyne or heterodyne detector as trusted loss and trusted noise. The trusted modelling can improve the system performance as it slightly restricts Eve's ability.

The EB version of the trusted model of a homodyne or heterodyne detector is depicted in Fig. 1. In this model, the limited detection efficiency is modeled by a beam splitter whose transmittance is used to imitate the detection efficiency, while the electronic noise is modeled by an Einstein-Podolsky-Rosen (EPR) source whose one mode is coupled into the quantum signal coming from the channel through the beam splitter, and the variance added by this coupling is used to imitate the electronic noise. Detailed analyses of the feasibility of conventional modeling are addressed in Ref.~\cite{Usenko_entropy_2009}.

\begin{figure}[t]
\centerline{\includegraphics[width=8.5cm]{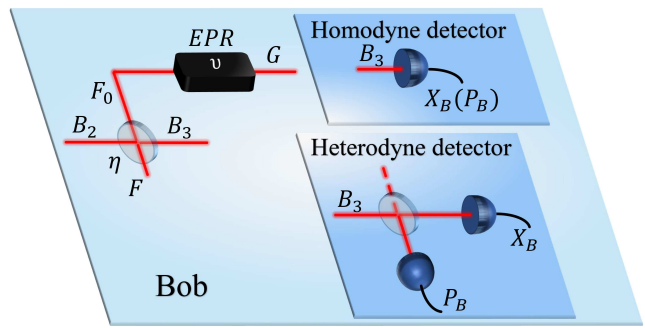}}
\caption{The detailed entangle-based trusted homodyne detector modelling, the transmittance of the beamsplitter is used to imitate the detection efficiency while the variance of the EPR state is used to imitate the electronic noise. The mode ${{B_3}}$ is then detected by either homodyne detecor or heterodyne detector.}\label{Fig1_conventional_7}
\end{figure}

The availability of both untrusted and trusted modeling relies on the equivalence of the EB model and the prepare-and- measure (PM) model. So, we write out the output of a practical homodyne detector in the corresponding PM model::

\begin{equation}
{X_{out}} = A{X_{LO}}\left( {\sqrt {{\eta _d}} {{\hat x}_B} + \sqrt {1 - {\eta _d}} {{\hat x}_{v1}}} \right) + {X_{ele}},
\end{equation}
where${X_{LO}}$ represents variable of the local oscillator, and ${X_{ele}}$ is a Gaussian variable with variance ${v_{el}}$ which normally has a Gaussian distribution, ${A}$ is the circuit amplification parameter.

However, sequence of the result from the homodyne detection can not be used directly. To the purpose of analyzing Eve'¡¯s information from the output ${X_{out}}$, we need to quantize this value using the SNU:
%in original model it is $SNU = {A^2}X_{LO}^2$. Then the data used for post-processing is:

\begin{equation}
x_{out}^{SNU} = \frac{{{X_{out}}}}{{\sqrt {SNU} }} = \left( {\sqrt {{\eta _d}} {{\hat x}_B} + \sqrt {1 - {\eta _d}} {{\hat x}_{v1}}} \right) + \frac{{{X_{ele}}}}{{A{X_{LO}}}}.
\end{equation}

The output sequence of ${x_{out}^{SNU}}$ corresponds to output of its EB model, which is depicted in FIG.1.

The idea of using a local local oscillator to interfere with the quantum signal in the detection stage can be used to develop defenses against all kinds of attack against the local oscillator. However, the local laser has fluctuations itself, and the electronic noise of the detector also suffers from fluctuations due to the changes in the environment, for example temperature changes. Thus, even when the LLO scheme is adopted, a calibration procedure is certainly required. The proposed measurement-device-independent protocols dedicated to defending against all kinds of attack against the homodyne detector also require proper calibration of the SNU~\cite{Li_QRA_2014, Zhang_QRA_2014, Pirandola_NatPhoton_2015}.

\subsection{Shot-noise unit calibration in the conventional approach}
In this subsection we review the conventional way of calibrating the SNU.

Typically, two step are required to perform the SNU calibration, \emph{Step 1}, one calculates the variance of the homodyne detecor output with both the quantum signal and the LO removed and takes this as the electronic noise ${V_{ele}}$. \emph{Step 2}, one calculates the variance of the homodyne detector output ${V_{tot}}$ with only the LO path connected, and takes this as the total noise, consisting of the electronic noise and the shot-noise unit. After these two steps, one can calculate the shot-noise unit through the equation:

\begin{equation}
SNU = {V_{tot}} - {V_{ele}}.
\end{equation}

In the following we also use ${SN{U^{TTE}}}$ to refer this kind of calibration method.
In practice, even under the assumption of untrusted modeling, the electronic noise is still known from the corresponding PM model regardless of the EB scenario adopted in the security analysis in order to obtain the SNU. Thus, trusted modeling is more reasonable in the EB scenario.

\subsection{Limitations with the conventional calibration}

First and foremost, the ${V_{tot}}$ in Eq. (3) is more than the vacuum noise plus the electronic noise, other noises including the relative-intensity noise (RIN) should be included.
So ${V_{tot}}$ should be rewritten as ${{V_{tot}} = SNU + {V_{ele}} + V{}_{RIN}}$, as is described in Eq. (3), the SNU calculated is ${SN{U^{TTE}} = SNU + V{}_{RIN}}$.
More precisely, the output of the practical homodyne detector should be rewrite as:
\begin{equation}
{X_{out}} = A{X_{LO}}\left( {\sqrt {{\eta _d}} {{\hat x}_B} + \sqrt {1 - {\eta _d}} {{\hat x}_{v1}}} \right) + {X_{ele} + X{}_{RIN}}.
\end{equation}

\begin{widetext}
The sequence of the raw key after the SNU normalization is thus given by:

\begin{equation}
x_{out}^{SN{U^{TTE}}} = \frac{{{X_{out}}}}{{\sqrt {SNU + {V_{RIN}}} }} = \frac{{A{X_{LO}}}}{{\sqrt {SNU + {V_{RIN}}} }}\left( {\sqrt {{\eta _d}} {{\hat x}_B} + \sqrt {1 - {\eta _d}} {{\hat x}_{v1}}} \right) + \frac{{{X_{ele}} + {X_{RIN}}}}{{\sqrt {SN{U} + {V_{RIN}}} }}.
\end{equation}

Thus, the equivalence between the PM model with the EB version of the model does not hold anymore.
\end{widetext}

Yet, as can be seen from Eq. (2), any incorrect estimated SNU will cause the false approximation of the statistics, furthermore, security analysis suggests that slightly error in the SNU estimation can greatly decrease the secret key rate.

\begin{figure}[t]
\centerline{\includegraphics[width=9.5cm]{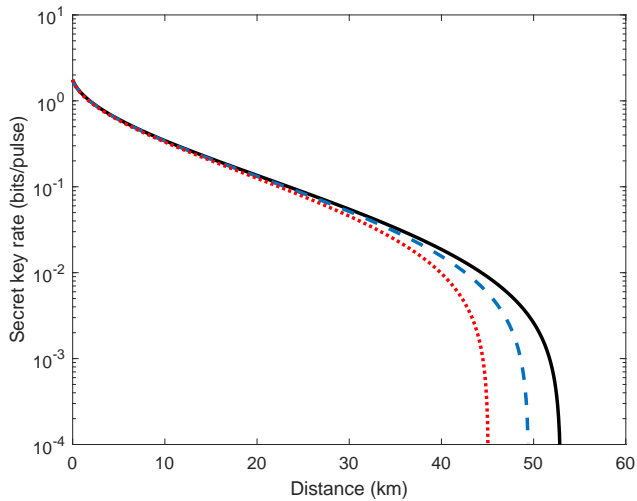}}
\caption{Simulation result of the secret key rate as a function of transmission distance with perfect SNU calibration as well as with error of 0.1\% and 0.3\% of SNU. The black solid line represents the secret key rate with a perfect SNU calibration. The blue dashed line represents the secret key rate with a 0.1\% of SNU calibration error whereas the red dotted line is the secret key rate with a 0.3\% of SNU calibration error.
The variance of the EPR is set to 40. The channel excess noise is in the simulation ${{\varepsilon _c} = 0.01}$, and the electronic noise ${{v_{ele}} = 0.01}$, the limited detection efficiency ${{\eta _d}{\rm{ = 0}}{\rm{.6}}}$ and the reconciliation efficiency ${\eta {\rm{ = 0}}{\rm{.956}}}$~\cite{Xiangyu_arxiv_2018}.
}
\end{figure}

In a realistic scenario, inaccuracies in the SNU can result in a drastic decrease in the secret key rate estimated through a security analysis. A simulation result is displayed in Fig. 2. It can be seen that even with a 0.1\% deviation from the real SNU, the secret key rate can drop significantly or even to zero when the transmission distance is over 50 km.

Moreover, attacks against the SNU can severely threat the security of the practical CV-QKD systems. For instance, the response curve of the homodyne detector is normally calibrated before the distribution stage, and so Eve can launch an attack that controls the LO signal~\cite{Jouguet_PhysRevA_2013}, which can delay the triggering of the detector. In this way, the actual slope of the response curve is decreased, and then if Bob still employs the original response curve to evaluate the SNU, the SNU will be overestimated, which leads to an underestimate of the excess noise. Also, an attack against the intensity of the LO~\cite{Ferenczi_OSA_2007} during the key distribution stage can also cause a misestimate of the SNU; the noise induced by Eve¡¯s attack may be underestimated if Bob applies the calibrated SNU to normalize his data. Thus a security loophole may be opened up.

In the conventional implementation, each optical path of the signal and LO requires an optical switch, which inevitably increases costs, and complicates the SNU calibration procedure and the data-processing procedures. Adding an optical switch in the LO path also decreases the optical power, which is not good for homodyne detection, which requires sufficient amplification of the LO signal.

Since the variance of the electronic noise and the total noise variance are measured separately, the SNU is not directly evaluated. According to Eq. (3), using subtraction to calculate the calibrated SNU introduces more statistical fluctuations since both the electronic noise and the total noise variance suffer from finite-size effects. In practice, we will need twice the calibration period to evaluate both ${V_{tot}}$ and ${V_{ele}}$, which suggests that the number of the data that are used to distill the key information also ineluctably reduced. As is discussed above, for a block length of $N$ samples, where $M_{1}$ samples are used to calculate the variance of the electronic noise, $M_{2}$ samples are used to compute the total noise variance, then only $N-M_{1}-M_{2}$ samples can be used to perform distillation. The above problems build barriers that will restrict the performance of the practical CV-QKD systems. In next section, we present the one-time-calibration model that is carefully designed to surmount the problems above.

\section{\label{sec:3} Calibration with one-time-calibration model}
In contrast to the original evaluation model, we now propose a calibration model which requires evaluation only one time. In this model, the limited detection efficiency and the electronic noise of the practical detector are still considered as trusted noise, and only one optical switch in the LO path is required in the corresponding PM scheme.

\subsection{Entanglement-based one-time-calibration model}
In this model, we still consider the electronic noise and the limited detection efficiency as the main imperfections of the practical homodyne detector.
Two beam splitters are applied to represent the electronic noise and the limited detection efficiency respectively in the corresponding EB model. In the following analysis, we will focus on the coherent state and homodyne detection scheme as an example. In this scenario, the electronic noise is modeled by the transmittance of the beam splitter. As is depicted in Fig. 3, the transmittance ${\eta _d}$ of the first beam splitter ${D_{1}}$ equals to the efficiency of the detector, while the transmittance ${\eta _e}$ of the second beam splitter ${D_{2}}$ models the electronic noise.

In order to keep consistent with the previous SNU calibration of the experimental demonstrations, in the following we still do not consider the RIN specifically, where we recognize the RIN is included in the ${SN{U^{TTE}}}$ of the conventional calibration model.
We only need to measure one time to identify the SNU when we exploit this model. The new SNU is measured as the output when the LO signal is on, so the new SNU can be rewritten as:

\begin{equation}
SNU' = {V_{tot}} = SNU+{V_{ele}},
\end{equation}
where ${SNU'}$ is the new calibrated shot-noise unit, ${SNU}$ is the shot-noise unit measured from two-time evaluation procedure.

With this model, certain advantages can be procured.
Firstly, we need only one optical switch in the signal path in our system. Secondly, calibration is required only one time in this model, which makes it a more useful model for application to practical systems. Thirdly, since we only need one-time period to calculate the shot-noise unit, the statistic fluctuation is minimized compared with original calibration model with two-time evaluation.

To analyze the one-time-calibration model in detail, we would like to restate the assumptions made about the homodyne detector in the first place:~\cite{Scarani_RevModPhys_2009, Pirandola_RevModPhys_2019}\\
1.The loss in Bob's side will not leak any information to Eve.\\
2.The electronic noise is not caused by Eve, and will not leak any information to Eve.\\
3.The electronic noise is additive Gaussian noise.

\begin{figure*}[t]
\centerline{\includegraphics[width=15.0cm]{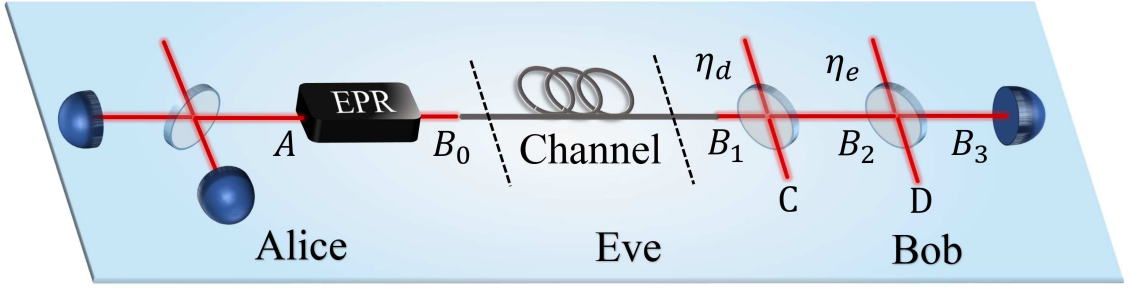}}
\caption{The entanglement-based model of one-time evaluation model. The model is based on coherent states and homodyne detection where Alice applies an heterodyne measurement on one mode of the EPR states, the other mode is sent to the quantum channel and measured by homodyne detection. The two other beamsplitters are used to imitate the electronic noise and the limited detection efficiency.}
\end{figure*}

It is also worth noting that in practical QKD implementations, quantum attacks against loopholes caused by imperfect devices may happen, which can cause practical security issues, but such attacks and the related defenses are not the emphasis of this paper. The analysis will still be restricted to the restatement above.
We choose the complete EB version of the one-time evaluation model for particular analysis. The beam splitter ${B1}$ is utilized to stimulate the electronic noise of the practical homodyne detector. Detailed analysis of the derivation are described as follows:
We start with considering the ideal homodyne detection, let us take x-quadrature for instance:

\begin{equation}
{\hat x_{\hom }} = \sqrt {{\eta _e}} \left( {\sqrt {{\eta _d}} {{\hat x}_B} + \sqrt {1 - {\eta _d}} {{\hat x}_{v1}}} \right) + \sqrt {1 - {\eta _e}} {\hat x_{v2}},
\end{equation}
where ${\hat x_{v1}},{\hat x_{v2}}$ indicate the vacuum coupled-in by the two beamsplitters. Vacuum state ${\hat x_{v1}},{\hat x_{v2}}$ are two Gaussian variables.

Nevertheless, we can consider the electronic noise as another ``optical mode'', and then we can define a joint shot-noise unit called $SN{U^{OTE}}$, which counts both the quantum signal and the ¡®optical mode¡¯. The new joint SNU is

\begin{equation}
SN{U^{OTE}} = {A^2}X_{LO}^2 + \left\langle {\Delta X_{ele}^2} \right\rangle  = {A^2}X_{LO}^2 + {v_{el}}.
\end{equation}

\begin{widetext}

So the data after quantization using new joint $SN{U^{OTE}}$ are

\begin{equation}
\chi _{out}^{SN{U^{OTE}}}= \frac{{{X_{out}}}}{{\sqrt {SN{U^{OTE}}} }} = \frac{{A{X_{LO}}}}{{\sqrt {{A^2}X_{LO}^2 + {v_{el}}} }}(\sqrt {{\eta _d}} {\hat x_B} + \sqrt {1 - {\eta _d}} {\hat x_{v1}}) + \frac{{{X_{ele}}}}{{\sqrt {{A^2}X_{LO}^2 + {v_{el}}} }}.
\end{equation}

Considering ${X_{ele}}$ is a Gaussian variable and it is immune to Eve, accordingly we can replace it with a Gaussian operator:$\sqrt {{v_{el}}} {\hat x_{v2}}$, in which ${\hat x_{v2}}$ has the variance of 1. Then

\begin{equation}
x_{out}^{SN{U^{OTE}}} = \frac{{A{X_{LO}}}}{{\sqrt {{A^2}X_{LO}^2 + {v_{el}}} }}\left( {\sqrt {{\eta _d}} {{\hat x}_B} + \sqrt {1 - {\eta _d}} {{\hat x}_{v1}}} \right) + \frac{{\sqrt {{v_{el}}} }}{{\sqrt {{A^2}X_{LO}^2 + {v_{el}}} }}{\hat x_{v2}}.
\end{equation}
\end{widetext}

Now if we define ${{\eta _e} = \frac{{{A^2}X_{LO}^2}}{{{A^2}X_{LO}^2 + {v_{el}}}}}$, then equation(8) can be rewrite as equation(3), then $x_{out}^{new} = {\hat x_{\hom }}$.

Thus we have also derived the corresponding EB version of the model and the equivalence between the PM model and the EB model is built. The PM model is used for actual implementation in the system while the EB model is used for security analysis. Our derivation guarantees that the measurement results of mode ${{B_3}}$ in the EB scheme are the same as the output of the PM scheme. Therefore, in this EB scheme, we never change the unit of the vacuum, the variance of the vacuum is always 1. This implies the EB model can describe the practical PM scheme where the measurement output is quantized by the ${SN{U^{OTE}}}$.

In this scheme, the dissipation caused by the imperfections of the homodyne detector can be observed intuitively. The electronic noise can be modeled as extra loss on the receiver¡¯s side. For example, we can assume that the clearance between the SNU and the electronic noise is normally required to be above 10 dB, and in total a 0.41 dB loss is acquired, which is equivalent to the loss in a 2.05 km fiber. For homodyne detectors that employ a higher clearance of the SNU over the electronic noise such as 15 dB, the total loss of the homodyne detector is around 0.132 dB, which is about 0.66 km of fiber loss, which is quite acceptable in a realistic setup.

Next we quickly review the case where the RIN is taking into the consideration:
In this case we identify the ${SN{U^{OTE}}}$ as
\begin{equation}
SN{U^{OTE}} = SNU + {V_{ele}} + {V_{RIN}}.
\end{equation}

To build the equivalence between the PM model and the corresponding EB model, we start by the output of the practical detector of the PM model in equation(1), next we quantize the output sequence of the detector by this ${SN{U^{OTE}}}$ as
\begin{widetext}
\begin{equation}
x_{out}^{} = \frac{{{X_{out}}}}{{\sqrt {SNU + {V_{ele}} + {V_{RIN}}} }} = \frac{{A{X_{LO}}}}{{\sqrt {SNU + {V_{ele}} + {V_{RIN}}} }}\left( {\sqrt {{\eta _d}} {{\hat x}_B} + \sqrt {1 - {\eta _d}} {{\hat x}_{v1}}} \right) + \frac{{{X_{ele}} + {X_{RIN}}}}{{\sqrt {SNU + {V_{ele}} + {V_{RIN}}} }}.
\end{equation}
\end{widetext}
In this scenario if we define ${\sqrt {{\eta _e}}  = \frac{{A{X_{LO}}}}{{\sqrt {SNU + {V_{ele}} + {V_{RIN}}} }}}$, then the output of the PM model is equal to that of the EB model in Fig. 3, and thus the equivalence between the EB model and the PM model holds, which proves that the one-time calibration model can extend the security analysis for trusted modeling and the SNU calibration process when any other additive noise is considered.

\subsection{One-time calibration method}

From the perspective of practical security, real-time calibration can defend against most attacks against the LO or detectors but requires seamless switching between SNU evaluation and key distribution. Thus one optical switch is needed in the signal path; when the optical switch is connected, the system performs key distribution, and when the optical switch is disconnected, the system performs SNU calibration.

With our intuition for solving problems presented in Sec.~\ref{sec:2}, we can circumvent usage of an optical switch in the LO path by adopting the one-time calibration process. This makes real-time calibration possible, since the optical switch on the LO side can decide whether the system is in the key distribution scheme or in the SNU calibration scheme. The optical power is retained maximally, and thus this also helps in amplifying the quantum signal.

\begin{figure}[b]
\centerline{\includegraphics[width=8.5cm]{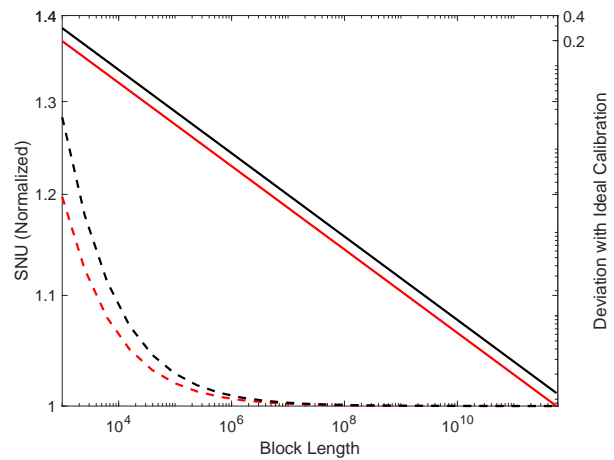}}
\caption{Normalized SNU and deviation with ideal calibration as a function of block length. Black lines are simulation results under the conventional model while red lines represent the one-time calibration model. Dash lines are simulation results of the normalized SNU as a function of the number of the calibrated data, solid lines are the deviation between the two models compare to ideal calibration. The total noise variance are set as 2.3768 where the electronic noise is 0.421. the failure probability of parameter estimation ${{\varepsilon _{PE}}}$, the failure probability during privacy amplification procedure ${{\varepsilon _{PA}}}$ and the smoothing parameter ${\bar \varepsilon }$ are all set as ${{10^{ - 5}}}$.}
\end{figure}

Furthermore, the data used to perform calibration are rather limited, especially when more data are required to distill key information. So, finite-size effects need to be carefully considered to help understand the effect of the finite data. For the one-time-calibration model the SNU in this scenario is directly measured thus the statistical fluctuation is suppressed. In Fig.4 we provide the simulation result that explores the SNU behaviour of the one-time calibration model as well as the conventional model.
In order to maximally imitate the practical realization we take the total noise variance as ${2.3768}$ and the electronic noise variance ${0.421}$. The simulation result shows the normalized SNU as a function of block length. In order to intuitively observe the deviation owing to the finite-size effects, we define another variable that measures the difference between the ideal calibration and the practical normalization. The simulation result is drawn in the right Y axis as a function of block length. Although the two model behave quite the same when the block length is over ${{10^8}}$, from the deviation compared to the ideal calibration
the one-time evaluation model has better performance than the conventional calibration model against finite-size effects.

\section{\label{sec:4} Performance analysis of one-time-calibration model}
The proposed one-time calibration model, which helps to reduce the system complexity and achieve seamless realtime SNU calibration, is distinctly more compliant with practical key distribution scenarios. Further, we extend the security analysis to a situation where the equivalence of the PM model and the EB model is built in the context of more practical implementations, where types of noise such as RIN can be taken into consideration. In this section, we give an exhaustive analysis of the performance of the one-time calibration model. Secret-key-rate calculations in both the asymptotic regime and the finite-size regime are provided, while numerous simulation results are illustrated to show their behavior compared with the original calibration model.

\subsection{Secret key rate calculation of one-time-calibration model}
The complete EB version of the one-time evaluation model is pictured in Fig.1. In our proposed model, the electronic noise ${\eta _e}$ is not directly measured, thereby the secret key rate calculation may not be so straightforward. As a preliminary consideration, Alice's data and Bob's data are directly acquirable from the experiment in the PM scheme thus mode A and mode B are what we can at least conclude in the corresponding EB model. Although mode ${C}$ and mode ${D}$ are not controlled by Eve, we merely ignore then at the moment. In this scenario the covariance matrix used to calculate the key rate is a four-order covariance matrix:

\begin{equation}
{\gamma _{A{B_3}}} = \left( {\begin{array}{*{20}{c}}
{{\gamma _A}}&{{\phi _{A{B_3}}}}\\
{\phi _{A{B_3}}^T}&{{\gamma _{{B_3}}}}
\end{array}} \right).
\end{equation}

The covariance matrix of $AB'$ can be deduced as:

\begin{equation}
{\gamma _{A{B_3}}} = \left( {\begin{array}{*{20}{c}}
{{V}{I_2}}&{\sqrt {{T}{\eta _d}{\eta _e}\left( {V^2 - 1} \right)} {\sigma _Z}}\\
{\sqrt {{T}{\eta _d}{\eta _e}\left( {V^2 - 1} \right)} {\sigma _Z}}&{\left[ {{T}{\eta _d}{\eta _e}\left( {{V} - 1 + {\varepsilon _c}} \right) + 1} \right]{I_2}}
\end{array}} \right),
\end{equation}
where ${V}$ is the variance of ${EPR}$ state in EB model, ${T}$ indicates channel transmissivity and ${{\varepsilon _c}}$ is the channel excess noise. ${I_2}$ is second-order identity matrix and ${\sigma _Z}$ is a 2*2 matrix: ${\left( {\begin{array}{*{20}{c}}
1&0\\
0&{ - 1}
\end{array}} \right)}$.

The main focus of secret-key-rate calculation is how to evaluate an upper bound on the information that Eve can procure.We restrict ourselves to reverse reconciliation~\cite{Devetak_ProcRSoc_2005}, and the secret key rate is calculated as:

\begin{equation}
R = \beta I(A:{B_3}) - I({B_3}:E),
\end{equation}
where ${\beta}$ is the reconciliation efficiency. The mutual information between the two legitimate parties Alice and Bob can be described by Shannon entropy which can be written as

\begin{equation}
{I_{A{B_3}}} = H({B_3}) - H({B_3}|A),
\end{equation}
where ${H(B)}$ can be calculated as ${\frac{1}{2}{\log _2}{V_B}}$, ${H(B|A)}$ is the conditional Shannon entropy that can be calculated as ${\frac{1}{2}{\log _2}{V_{B|A}}}$, the conditional variance ${V_{B|A}}$ means the remaining uncertainty on Bob's variance after the measurement on Alice's side. According to equation(10), we can derive ${I_{AB}}$ as

\begin{equation}
{I_{A{B_3}}} = \frac{1}{2}{\log _2}(\frac{{{V} + \chi }}{{\chi  + 1}}),
\end{equation}
where we define ${\chi}$ as ${\chi  = \frac{1}{{T{\eta _d}{\eta _e}}} - 1 + {\varepsilon _c}}$ for conciseness.

Now we derive the mutual information between Bob and Eve, the maximum information between Eve and Bob is decided by the Holevo interval~\cite{Nielsen_QCQI} ${{\chi _{{B_3}E}}}$,

\begin{equation}
{\chi _{{B_3}E}} = S({\rho _E}) - \int {dm{}_{B_3}p({m_{B_3}})S(\rho _E^{{m_{B_3}}})} .
\end{equation}

According to the Fig. 3, the eavesdropper Eve and the measurement will both purify the system ${AB'}$, then ${{\chi _{BE}}}$ should be written as

\begin{equation}
{\chi _{{B_3}E}} = S({\rho _{A{B_3}}}) - S(\rho _A^{{m_{{B_3}}}}),
\end{equation}
where ${S({\rho _{A{B_3}}})}$ and ${S(\rho _A^{{m_{{B_3}}}}})$ can be figured out by calculating the symplectic eigenvalues of the correspondence covariance matrix ${{\gamma _{A{B_1}}}}$ and ${\gamma _A^{{m_{B_3}}}}$, where ${\gamma _A^{{m_{B_3}}}}$ is the covariance matrix of modes ${A}$ and ${{B_3}}$ after Bob performing detections:

\begin{equation}
{\chi _{{B_3}E}} = \sum\nolimits_{i = 1}^2 {G(\frac{{{\lambda _i} - 1}}{2}) - } \sum\nolimits_{i = 3}^4 {G(\frac{{{\lambda _i} - 1}}{2}),}
\end{equation}
where ${G(x) = (x + 1){\log _2}(x + 1)-x{\log_2}x}$, and ${{\lambda _i}}$ are the symplectic eigenvalues.
We have already derived the covariance matrix of ${A{B_1}}$, so its symplectic eigenvalues ${\lambda _1}$, ${\lambda _2}$ can be subsequently derived:

\begin{equation}
\begin{array}{l}
\lambda _{1,2}^2 = \frac{1}{2}[A \pm \sqrt {{A^2} - 4B} ],\\
A = {{V}^2}(1 - 2T{\eta _e}{\eta _d}) + 2T{\eta _e}{\eta _d} + {(T{\eta _e}{\eta _d})^2}{({V} + \chi )^2},\\
B = {(T{\eta _e}{\eta _d})^2}{({V}\chi  + 1)^2}.
\end{array}
\end{equation}

Next we calculate the symplectic eigenvalues of the covariance matrix of ${\gamma _A^{{m_B}}}$. After the quantum signals arrive at Bob's side, Bob can take whether homodyne detection or heterodyne detection to measure the quantum states. Considering most practical systems would use homodyne detection at present, in this paper we analyze the sympletic eigenvalues after Bob performs homodyne detection.

Bob performs homodyne detection on mode ${B¡¯}$, measuring whether its coordinate $x$ or momentum $p$. After detection, mode A will be projected to new Gaussian state, where the covariance matrix will become:

\begin{equation}
\gamma _A^{{m_{B_3}}} = {\gamma _A} - {\sigma _{A{B_3}}}{(X{\gamma _{B_3}}X)^{MP}}\sigma _{A{B_3}}^T,
\end{equation}
where ${\gamma _A}$, ${\gamma _{B_3}}$, ${\sigma _{A{B_3}}}$ are corresponded with equation(9), and we find that ${\gamma _A^{{m_{B_3}}}}$ takes the form:

\begin{equation}
\gamma _A^{{m_{B_3}}} = \left( {\begin{array}{*{20}{c}}
{\frac{{{V}\chi  + 1}}{{{V} + \chi }}}&0\\
0&{V}
\end{array}} \right).
\end{equation}

The symplectic eigenvalue of this matrix is:

\begin{equation}
\lambda _3^2 = {V}(\frac{{{V}\chi  + 1}}{{{V} + {V}\chi }}).
\end{equation}

By combining this with Eqs. (17) and (20), the secret key rate is calculable.

In the above scenario we include mode ${A}$ and mode ${B_3}$ into our security analysis, we call this the two-mode EB model. However, we scarcely exploit the features of the two beam splitters that represent the homodyne  detector imperfections in the EB model which may result in the underestimating of the secret key rate. In the following we manage to make these imperfections into trusted losses. By permuting the two beam splitters, we can take the mode ${C}$ that models the detector efficiency into Bob's side.
The feasibility of the permuting operation is based on the trusted-homodyne-detector assumption. The eavesdropper Eve cannot take control of the homodyne detector, the permuting operation will not affect detected mode ${{B_3}}$, thus it will not influence Eve's knowledge about Bob's result. After the permuting of the beams plitters, we obtain a new EB model and since mode ${{B_3}}$ remains unchanged, this new EB scheme is still a match with the PM scheme that uses ${SN{U^{OTE}}}$ to quantise the measurement output.
The complete EB version of this scenario is traced out in Fig. 5 and in total modes ${A}$, ${{B_3}}$ and ${C}$ are comprised into consideration.
In this case mode ${D}$ that represents the electronic noise is still unknown to us,  but the detection efficiency is a known quantity. Thus we may obtain the covariance matrix with there mode ${A}$, ${{B_3}}$ and ${C}$, we call it three-mode EB model, and it should take the form

\begin{equation}
{\gamma _{AC{B_3}}} = \left( {\begin{array}{*{20}{c}}
{{\gamma _A}}&{{\phi _{AC}}}&{{\phi _{A{B_3}}}}\\
{\phi _{AC}^T}&{{\gamma _C}}&{{\phi _{C{B_3}}}}\\
{\phi _{A{B_3}}^T}&{\phi _{C{B_3}}^T}&{{\gamma _{{B_3}}}}
\end{array}} \right).
\end{equation}

\begin{figure*}[t]
\centerline{\includegraphics[width=15cm]{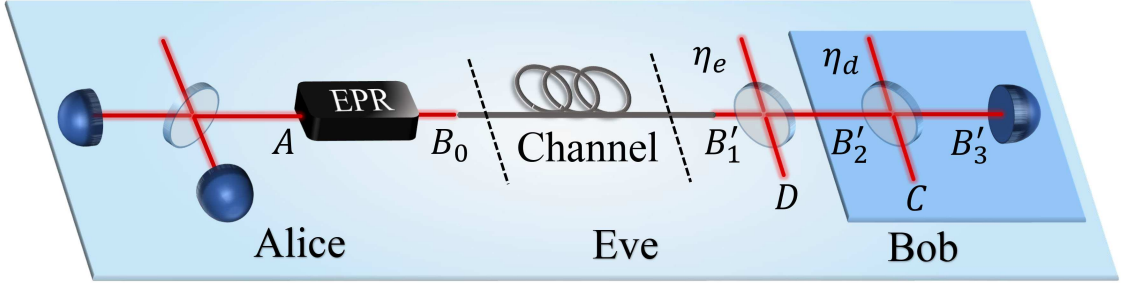}}
\caption{Three-mode entanglement-based model of the one-time-calibration model. Coherent states and homodyne scenario are taken as an example, where Alice one mode of the EPR source is measured by the heterodyne detection while the  other mode is sent through the quantum channel and measured by Bob using homodyne detection. In this three-mode EB scenario we treat the limited detection efficiency as a trusted loss thus it is been taken into Bob's side. While the electronic noise has been considered as a channel loss.}\label{picx1}
\end{figure*}

Alice first generate a two-mode squeezed state then send one of its mode to the quantum channel, before it goes through the first BS, the state of the system ${{A{B_1}^\prime }}$ is a pure state with the expectation of 0, the covariance matrix of modes ${A}$ and ${{B_1}^\prime }$ is

\begin{equation}
{\gamma _{A{B_1}^\prime }} = \left( {\begin{array}{*{20}{c}}
{{V}{I_2}}&{\sqrt {T(V^2 - 1)} {\sigma _z}}\\
{\sqrt {T(V^2 - 1)} {\sigma _z}}&{[T({V} - 1 + {\varepsilon _c}) + 1]{I_2}}
\end{array}} \right).
\end{equation}

After the quantum signal goes through the first BS, ${A{B_2}^\prime D}$ is a pure state, it can be described with covariance matrix ${\gamma _{A{B_2}^\prime D}}$

\begin{equation}
{\gamma _{A{B_2}^\prime D}} = {({Y^{BS}})^T}[{\gamma _{A{B_1}^\prime}} \oplus {I_2}]({Y^{BS}}).
\end{equation}

The covariance matrix ${{\gamma _{A{B_1}^\prime}}}$ performs kronecker product with vacuum state since no other signal is needed to couple in the beamsplitter in this scenario. ${{Y^{BS}}}$ is the transformation matrix of the first beamsplitter, which models the electronic noise of the practical homodyne detector:

\begin{equation}
{Y^{BS}} = {I_2} \oplus Y_{{\eta _e}}^{BS},
\end{equation}
where ${Y_{^{{\eta _e}}}^{BS}}$ is the symplectic matrix of the beam splitter

\begin{equation}
Y_{{\eta _e}}^{BS} = \left( {\begin{array}{*{20}{c}}
{\sqrt {{\eta _e}} {I_2}}&{\sqrt {1 - {\eta _e}} {I_2}}\\
{ - \sqrt {1 - {\eta _e}} {I_2}}&{\sqrt {{\eta _e}} {I_2}}
\end{array}} \right).
\end{equation}

However we do not measure the electronic noise directly which implies we do not know the mode ${D}$ in this one-time evolution model. For the rest, the covariance matrix ${{\gamma _{A{B_2}^\prime}}}$ is

\begin{equation}
{\gamma _{A{B_2}^\prime}} = \left( {\begin{array}{*{20}{c}}
{{V}{I_2}}&{\sqrt {T{\eta _e}(V^2 - 1)} {\sigma _z}}\\
{\sqrt {T{\eta _e}(V^2 - 1)} {\sigma _z}}&{[T{\eta _e}({V} - 1 + {\varepsilon _c}) + 1]{I_2}}
\end{array}} \right).
\end{equation}

After passes through the second beamsplitter, which is considered as a trusted loss, the covariance matrix ${{\gamma _{A{B_3}C}}}$ should be able to procure:

\begin{equation}
{\gamma _{A{B_3}C}} = {({Y^{BS'}})^T}[{\gamma _{A{B_2}}} \oplus {I_2}]({Y^{BS'}}),
\end{equation}
where ${{Y^{BS'}}}$ is the transformation matrix that imitate the limited detection efficiency of the practical beam splitter:

\begin{equation}
{Y^{BS'}} = I{I_2} \oplus Y_{{\eta _d}}^{BS}.
\end{equation}

${Y_{^{{\eta _d}}}^{BS}}$ is the symplectic matrix of the beamsplitter:

\begin{equation}
Y_{{\eta _d}}^{BS} = \left( {\begin{array}{*{20}{c}}
{\sqrt {{\eta _d}} {I_2}}&{\sqrt {1 - {\eta _d}} {I_2}}\\
{ - \sqrt {1 - {\eta _d}} {I_2}}&{\sqrt {{\eta _d}} {I_2}}
\end{array}} \right).
\end{equation}

Now the covariance matrix of three modes ${A, {B_3}}$ and ${C}$ can be calculated using Eq. (31). And it is supposed to achieve a higher secret key rate compare to the two-modes model mentioned earlier:

\begin{widetext}
\begin{equation}
{\gamma _{A{B_3}C}} = \left( {\begin{array}{*{20}{c}}
{V{I_2}}&{\sqrt {T{\eta _e}{\eta _d}({V^2} - 1)} {\sigma _z}}&{\sqrt {T{\eta _e}(1 - {\eta _d})({V^2} - 1)} {\sigma _z}}\\
{\sqrt {T{\eta _e}{\eta _d}({V^2} - 1)} {\sigma _z}}&{[T{\eta _e}{\eta _d}(V - 1 + {\varepsilon _c}) + 1]{I_2}}&{\sqrt {{\eta _d}(1 - {\eta _d})} T{\eta _e}(V - 1 + {\varepsilon _c}){I_2}}\\
{\sqrt {T{\eta _e}(1 - {\eta _d})({V^2} - 1)} {\sigma _z}}&{\sqrt {{\eta _d}(1 - {\eta _d})} T{\eta _e}(V - 1 + {\varepsilon _c}){I_2}}&{[T{\eta _e}(1 - {\eta _d})(V - 1 + {\varepsilon _c}) + 1]{I_2}}
\end{array}} \right).
\end{equation}
\end{widetext}

In this case, the secret key rate is calculated in the same way as in Eq. (15). The mutual information between Alice and Bob is calculated from Eq. (16), with the variance of ${EPR}$ state at Alice's side ${{V}}$, variance at Bob's side ${(T{\eta _e}{\eta _d}({V} - 1 + {\varepsilon _c}) + 1)}$ and covariance between Alice and Bob ${(T{\eta _e}{\eta _d}(V^2 - 1))}$ which gives exactly the same results as the two-mode model scenario as Eq. (17).

Next we estimate the upper bound of information between Eve and Bob, according to Eq. (18), ${{\chi _{BE}}}$ is now rewritten as:

\begin{equation}
{\chi _{BE}} = S({\rho _{A{B_3}C}}) - S(\rho _{AC}^{{m_{{B_3}}}}).
\end{equation}

${S({\rho _{A{B_3}C}})}$ is Eve's Von Neumann entropy, since Eve purifies the system ${{A{B_3}C}}$. It can be figured out with the symplectic eigenvalues of the corresponding covariance matrix ${{\gamma _{A{B_3}C}}}$, but there is one eigenvalue that constantly equals to one, which means it contains no information, while the other symplectic eigenvalues contributes to the entropy. ${S(\rho _{AC}^{{m_{{B_3}}}})}$ is the Von Neumann entropy of the remaining quantum states have after Bob performs homodyne detection. So over all, the mutual information between Eve and Bob can be can be further exploited as in Eq. (20).

${{\lambda _1}}$ and ${{\lambda _2}}$ are derived from covariance matrix ${S({\rho _{A{B_3}C}})}$:

\begin{equation}
\lambda _{1,2}^2 = \frac{1}{2}[A \pm \sqrt {{A^2} - 4B} ],
\end{equation}
where parameter ${A}$ and ${B}$ are:

%\begin{multline}
%\begin{equation}
\begin{eqnarray}
\begin{array}{l}
 A = C(2 + C) - D + 1, \\
B = {V^2}{[(1 - {\eta _d}){\eta _d}{C^2} + C + 1]^2} -  \\
{V}{[(1 - {\eta _d}){\eta _d}]^2}{C^2}D - (1 - {\eta _d}){\eta _d}[(1 - {\eta _d}){\eta _d}{C^2} + C + 1]{D^2}.
\end{array}
\end{eqnarray}

Here we note ${C = T{\eta _e}(V - 1 + {\varepsilon _c})}$, ${D = T{\eta _e}(V^2 - 1)}$ for shortness.

Next we figure out the symplectic eigenvalues of matrix ${\gamma _{AC}^{{m_{{B_3}}}}}$ to calculate ${S(\rho _{AC}^{{m_{{B_3}}}})}$. ${\gamma _{AC}^{{m_{{B_3}}}}}$ is the covariance matrix after Bob applies homodyne detection which can be derived from:

\begin{equation}
\gamma _{AC}^{{m_{{B_3}}}} = {\gamma _{AC}} - {\phi _{AC{B_3}}}{(X{\gamma _{{B_3}}}X)^{MP}}\phi _{AC{B_3}}^T.
\end{equation}

It has two non-zero symplectic eigenvalues so the symplectic eigenvalues should have form of:

\begin{equation}
\lambda _{3,\;4}^2{\rm{ = }}\frac{{\rm{1}}}{{\rm{2}}}[{E^2} \pm \sqrt {{E^2} - 4{F^2}} ],
\end{equation}
where ${E}$, ${F}$ are defined as:
\begin{eqnarray}
\begin{array}{l}
E = \frac{{GV + {V^2} - 2(1 - {\eta _d})D + (C + 1)[(1 - {\eta _d})C + 1]}}{{{\eta _d}C + 1}}, \\
F = [\frac{{{\eta _d}G + V}}{{{\eta _d}C}} - (1 - {\eta _d})D{(\frac{{{\eta _d}C}}{{{\eta _d}C + 1}} - 1)^2}][(1 - {\eta _d})G + V],
\end{array}
\end{eqnarray}
where ${C}$ and ${D}$ have been defined previously. We noted symbol ${G}$ as ${G = T{\eta _e}[V({\varepsilon _c} - 1) + 1]}$ for shortness.

However, we are still not satisfy that mode $D$ is practically controlled by Eve, as we do not know the electronic noise at this stage. The electronic noise is viewed as the transmittance of a beam splitter, which can be treated as a loss. In realistic experiments, this loss may be calculated from Alice and Bob¡¯s data; it can be expressed as the product of the channel transmittance ${T}$ and the transmittance of the beamsplitter ${\eta {}_e}$. To the purpose of ultimately finding out the value of ${\eta {}_e}$, we can find some restrictions to limit the possible values of ${\eta {}_e}$:

\begin{eqnarray}
\left\{ {\begin{array}{*{20}{c}}
{const. \le T \le 1,}\\
{const. \le {\eta _e} \le 1,}\\
{T{\eta _e} = const,}
\end{array}} \right.
\end{eqnarray}
where we assume that $const.$ is the transmittance that represents the total loss from channel and electronic noise. Although it seems that there are plenty of valid values that ${\eta {}_e}$ may take, we still have to limit ourselves so that the lowest secret key rate can be achieved. And we note that the value is when the channel transmittance ${T}$ takes the value of $const.$, thereafter ${\eta {}_e}$ will may only take the value of 1. It suggests that there is no electronic noise existed, all that loss is contributed by the untrusted channel, essentially, this resultant is realised because the lowest secret key rate may be obtained when the untrusted party controls all the loss. It will no longer have mode ${D}$ and the EB model virtually retrogrades to the scenario where we only consider ${A, {B_3}}$ and ${C}$ aforementioned. Therefore, we may not consider Eve purifies ${A, {B_3}, D}$ and ${C}$ as ${S(\rho {}_E) = S({\rho _{ACD{B_3}}})}$ in the EB model, but instead she purifies them as ${S(\rho {}_E) = S({\rho _{AC{B_3}}}),}$ in the form of equation (35).

\subsection{Application in CV-QKD system finite-size regime}
In this section, we analyze the finite-size regime on the proposed models as well as the original model and mainly focus on the effect on shot-noise calibration. We first analyze its influence on shot-noise calibration procedure, then calculate the secret key rate of the proposed models based on covariance matrices.

As is discussed in section \uppercase\expandafter{\romannumeral3}, the original calibration procedure needs two steps to obtain the SNU result.
Regarding the new model however, we define the measurement results of the output when the LO path is connected as shot-noise unit, as in Eq. (3). According to the theory of finite-size analysis~\cite{Leverrier_PhysRevA_2010}, a finite-size block will introduce more statistical fluctuations compared with the asymptotic regime, which will lead to a decrease in the secret key rate. The importance of conducting finite-size analysis of CV-QKD systems has been extensively demonstrated. Consequently, the secret-key-rate calculation needs to be revised

\begin{equation}
R = \frac{n}{N}[\beta I(A:B) - {I_{{\varepsilon _{PE}}}}(B:E) - \Delta (n)],
\end{equation}
where ${N}$ is the block length and ${n}$ is the number of data that used to distill the key information.

We now consider the influence on shot-noise unit of finite-size regime. Maximum likelihood estimation is applied and we can describe the shot-noise unit as

\begin{equation}
{\hat V_{tot}} = \frac{1}{m}\sum\nolimits_{i = 1}^m {{{({y_{0i}})}^2}} .
\end{equation}

The 'hat' in the above equation suggests that it is an estimated value. We then apply law of large numbers to acquire the approximate distribution of the shot-noise unit

\begin{equation}
\frac{{m{{\hat V}_{tot}}}}{{{V_{tot}}}} \sim {\chi ^2}(m - 1),
\end{equation}
where ${V_{tot}}$ is the actual value of the variance of the SNU, ${{\chi ^2}}$ means chi-square distribution. By combining this with the failure probability of the parameter estimation process ${{\varepsilon _{PE}}}$, we can compute the fluctuation interval for an identified confidence interval.We can seek out the worst case. Here, we set the confidence interval as ${{\varepsilon _{PE}}/2}$, so the range of the fluctuated shot-noise unit can be calculated as

\begin{equation}
\Delta {V_{tot}} = {{z_{{\varepsilon _{PE/2}}}}}\frac{{{{\hat V}_{tot}}\sqrt 2 }}{{\sqrt m }},
\end{equation}
where ${{z_{\varepsilon PE/2}}}$ satisfies ${1 - erf({z_{\varepsilon PE/2}}/\sqrt 2 )/2 = \varepsilon PE/2}$, and ${erf}$ is the error function that defined as

\begin{equation}
erf(x) = \frac{2}{{\sqrt \pi  }}\int_0^x {{e^{ - {t^2}}}dt} .
\end{equation}

Therefore, the final shot-noise unit ${SNU'}$ is
\begin{equation}
SNU' \in [{\hat V_{tot}} - \Delta {V_{tot}},\;{\hat V_{tot}} + \Delta {V_{tot}}].
\end{equation}

Based on the explication above, the confidence interval of shot-noise unit of the original two-time evaluation model under finite-size effect should be described a
\begin{equation}
SNU = [{\hat V_{tot}} - \Delta {V_{tot}} - {\hat V_{ele}} - \Delta {V_{ele}},\;{\hat V_{tot}} + \Delta {V_{tot}} - {\hat V_{ele}} + \Delta {V_{ele}}].
\end{equation}

${{\hat V_{ele}}}$ is the estimated electronic noise which is corresponding to the system output when both the quantum signal path and LO path are turned off. And consequently ${\Delta {V_{ele}}}$ is the statistical fluctuation that this measurement brings in.

Since the influence of finite-size effect on shot-noise unit has already been studied in detail, next we exploit how it changes the secret-key-rate calculation. We noted the shot-noise-unit as ${N_{0}}$ for simplicity. For the two-mode model scenario, the variance of mode ${B_{3}}$ and covariance ${AB_{3}}$ vary as the shot-noise unit changes. So the covariance matrix needs to be rewritten in order to take consideration of the effect of shot-noise unit

\begin{widetext}
\begin{equation}
{\gamma _{A{B_3}}} = \left( {\begin{array}{*{20}{c}}
{{V}{I_2}}&{\sqrt {T{\eta _e}{\eta _d}(V^2 - 1)/{N_0}} {\sigma _z}}\\
{\sqrt {T{\eta _e}{\eta _d}(V^2 - 1)/{N_0}} {\sigma _z}}&{\{ [T{\eta _e}{\eta _d}({V} - 1 + {\varepsilon _c}) + 1]/{N_0}\} {I_2}}
\end{array}} \right).
\end{equation}
\end{widetext}

With the derived covariance matrix ${\gamma _{A{B_3}}}$, the following procedure for the secret-key-rate calculation is quite similar to the case in the asymptotic-limit regime, and, by traversing through the confidence interval, a lower bound on the secret key rate with finite-size effects can be found.

The derivation of the three-mode covariance matrix is little complicated. We need to reconsider the elements of the covariance matrix which vary because of the shotnoise- unit fluctuations. As described above, the variance for Bob is modified to ${[T{\eta _e}{\eta _d}({V} - 1 + {\varepsilon _c}) + 1]/{N_0}}$, the variance of mode ${C}$ can be modified by using the element ${{V_{{B_3}}}}$ as follows

\begin{equation}
{V_C} = \frac{{({V_{{B_3}}} - 1)}}{{{\eta _d}}}(1 - {\eta _d}) + 1.
\end{equation}

Further the covariance elements of the matrix also alter as
\begin{equation}
\begin{array}{l}
 < A{B_3} >  = \sqrt {T{\eta _d}{\eta _e}({V^2} - 1)/{N_0}} ,\\
 < AC >  =  -  < A{B_3} > \sqrt {(1 - {\eta _d})/{\eta _d}} ,\\
 < {B_3}C >  =  - ({V_{{B_3}}} - 1)\sqrt {(1 - {\eta _d}){\eta _d}} /{\eta _d}.\\
\end{array}
\end{equation}

So the elements in covariance matrix ${{\gamma _{AC{B_3}}}}$ which the shot-noise unit attributes to have all been modified and the matrix is rewritten as

\begin{widetext}
\begin{equation}
{\gamma _{AC{B_3}}} = \left( {\begin{array}{*{20}{c}}
{{V}{I_2}}&{ - \sqrt {T{\eta _e}(1 - {\eta _d})(V^2 - 1)/{N_0}} {\sigma _z}}&{\sqrt {T{\eta _e}{\eta _d}(V^2 - 1)/{N_0}} {\sigma _z}}\\
{ - \sqrt {T{\eta _e}(1 - {\eta _d})(V^2 - 1)/{N_0}} {\sigma _z}}&{[({V_{{B_3}}} - 1)(1 - {\eta _d})/{\eta _d}]{I_2}}&{ - [({V_{{B_3}}} - 1)\sqrt {{\eta _d}(1 - {\eta _d})} /{\eta _d}]{\sigma _z}}\\
{\sqrt {T{\eta _e}{\eta _d}(V^2 - 1)/{N_0}} {\sigma _z}}&{ - [({V_{{B_3}}} - 1)\sqrt {{\eta _d}(1 - {\eta _d})} /{\eta _d}]{\sigma _z}}&{{V_{{B_3}}}{I_2}}
\end{array}} \right).
\end{equation}
\end{widetext}

By setting the failure probability of the parameter estimation, the confidence interval can be identified. Then a lower bound on the secret key rate with finite-size effects can be procured by traversing the value of the shot-noise unit through the confidence interval. With the modified covariance matrix ${{\gamma _{AC{B_3}}}}$, further procedures to calculate secret key rate can be followed.

\subsection{Numerical simulation and discussion}
In this subsection, we give numerical simulation results of firstly the two one-time-calibration model aforementioned as well as the original two-time calibration model to make comparison. The comparisons between these models under finite-size regime are also provided subsequently.

\begin{figure}[b]
\centerline{\includegraphics[width=9.5cm]{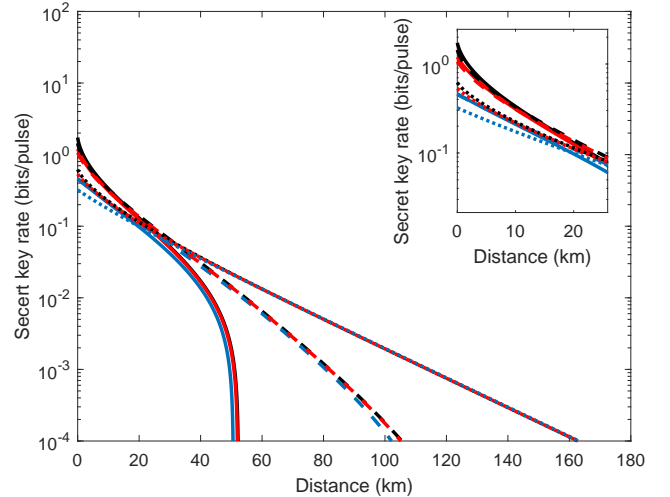}}
\caption{Secret key rate as a function of transmission distance with the original two-time-evaluation model and two one-time-evaluation models under different variances in the asymmetric scenario in case of the distance from 0 to 200km(main figure) and in case of the distance from 0 to 25km(inset). The channel excess noise is set as ${{\varepsilon _c} = 0.01}$, the electronic noise ${{v_{ele}} = 0.01}$, the limited detection efficiency ${\eta_{d}=0.6}$ and the reconciliation efficiency ${\eta  = 0.956}$.}\label{simu2andAmplify}
\end{figure}

Fig.6 shows the secret key rate as a function of the transmission distance for different SNU calibration models. The solid line, dashed line and pointed line represent simulation results under different variance ${V}$ of 40, 20 and 4 respectively. The red lines represent the three-mode one-time calibration model, the blue lines represent the two-mode one-time calibration model, and the black lines represents original evaluation model. The secret key rates among these three models can be exceedingly close under variance ${V=4}$, for the scenarios of variance ${V=20}$ and ${V=40}$, the secret key rate of three-mode one-time-calibration model and original evaluation model are still very close while the secret key rate of the two-mode model is slightly lower than the other two models. This implies that the one-time evaluation models we proposed are suitable in estimating the secret key rate especially for the three-mode one-time evaluation model, which is considerably approaching to the original two-time evaluation model.

Fig.7 is the simulation result of tolerable excess noise (TEN) versus the transmission distance: the solid line, dashed line and pointed line are results under variance ${V}$ of 40, 20 and 4 respectively. The red lines denotes the three-mode one-time-calibration model, the blue lines denotes two-mode one-time-calibration model and the black lines denotes the original evaluation model. It can be seen that three-mode model and original two-time-evaluation model can tolerate the highest excess noise under all three different variances whereas the two-mode one-time-evaluation model is not as outperformance as the other two models.

\begin{figure}[t]
%\centerline{\includegraphics[width=9.5cm]{simu3excess81.eps}}
\centerline{\includegraphics[width=9.5cm]{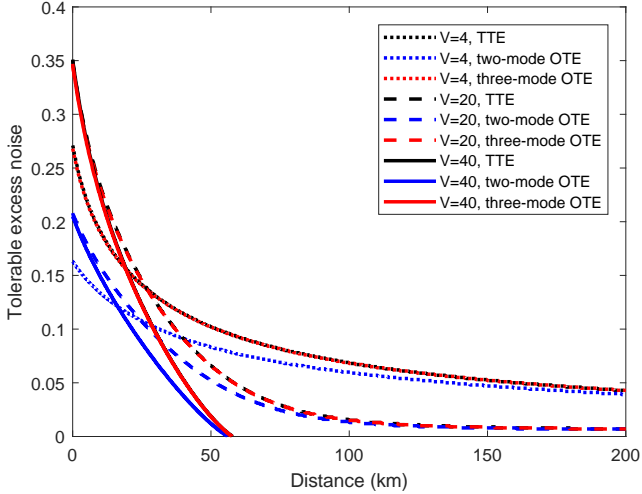}}
\caption{Tolerable excess noise as a function of transmission distance with the original two-time-evaluation model and the two one-time-evaluation models under different variances. The electronic noise ${{v_{ele}} = 0.01}$, the limited detection efficiency ${{\eta _d} = 0.6}$ and the reconciliation efficiency ${\eta  = 0.956}$.}\label{simu3excess81}
\end{figure}

We define a secret-key-rate disparity as ${\frac{{\left| {{R_{new}} - {R_{original}}} \right|}}{{{R_{original}}}}}$ to intuitively demonstrate the degree of difference between the one-time-calibration models with the original two-time-calibration model. From the previous simulation results we realise that the secret-key-rates of the one-time-calibration models are not exceeding that of the two-time-calibration model. Thus this disparity can show how close the two one-time-calibration models approach to the two-time-calibration model.
\begin{figure}[t]
\centerline{\includegraphics[width=9.5cm]{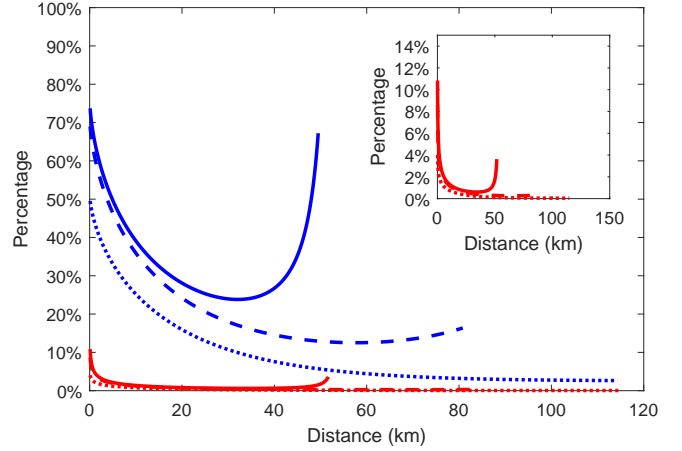}}
\caption{Secret key rate disparity as a function of transmission distance with two one-time-evaluation models over original two-time-evaluation model under different variances. Channel excess noise is set as ${{\varepsilon _c} = 0.01}$, the electronic noise ${{v_{ele}} = 0.01}$, the limited detection efficiency ${{\eta _d} = 0.6}$ and the reconciliation efficiency ${\eta  = 0.956}$.}\label{simuAsympto5}
\end{figure}

A lower percentage suggests a smaller divergence in the secret key rate compares to the two-time-calibration model. As can be seen from Fig. 8, the three-mode model is considerably outperformance the two-mode model. The secret key rate of the three-mode model with variance 4 can achieve a as low as 0.64\% divergence compare to the two-time-calibration model.
On balance, the one-time evaluation model is rather appropriate for estimating a lower bound on the information that Alice and Bob can share, and the one-time calibration model with three modes performs better than the two-mode one-time evaluation model.

Both Fig.9 and Fig.10 show the protocol performance of the two one-time-calibration models and the original two-time-evaluation model under the finite-size regime. The following parameters are taken into the simulations: the failure probability of parameter estimation ${{\varepsilon _{PE}}}$, the failure probability during privacy amplification procedure ${{\varepsilon _{PA}}}$ and the smoothing parameter ${\bar \varepsilon }$ are all set as ${{10^{ - 10}}}$~\cite{Leverrier_PhysRevA_2010}. The dimension of the Hilbert space of the variable ${x}$ in the raw key is set as ${\dim {H_x} = 2}$. The block length is set as ${{10^{ 10}}}$ and and half of the data are used to perform parameter evaluation, and so the remainder is used to extract the secret key. The one-time calibration model in the two-mode scenario is depicted using blue lines, the three-mode scenario is depicted using red lines, and the while black lines represent the original two-time evaluation model. The solid, dashed, and dotted lines correspond to variances ${{V} = 40, 20, 4}$ respective.

\begin{figure}[t]
\centerline{\includegraphics[width=9.5cm]{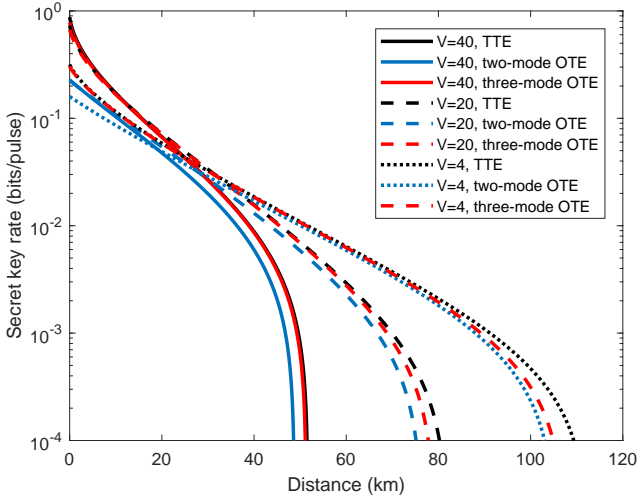}}
\caption{Secret key rate as a function of transmission distance with the original two-time-evaluation model and the two one-time-evaluation models under different variances ${{V} = 40, 20, 4}$ in the finite-size regime.
Electronic noise is set as ${{v_{ele}} = 0.01}$, the channel excess noise ${{\varepsilon _c} = 0.01}$, the limited detection efficiency ${{\eta _d} = 0.6}$ and the reconciliation efficiency ${\eta  = 0.956}$.}\label{FiniteSize_KR_4682}
\end{figure}

Fig.9 shows the secret key rate as a function of transmission distance, solid line, dashed line and pointed line are results under variance ${V}$ of 40, 20 and 4 respectively. The red lines denotes the three-mode one-time-calibration model, the blue lines denotes two-mode one-time-calibration model and the black lines denotes the original evaluation model.
In the case of variance ${{V} = 40}$ the secret key rates of both two-mode one-time-calibration model and three-mode one-time-calibration model are as quite close to the original two-time-evaluation model for about 70km, after that, the original evaluation model achieves a sightly higher secret key rate. For a smaller variance of ${{V} = 4}$, the secret key rate of three-mode one-time-calibration model is basically the same as the original two-time-evaluation model, also, the transmission distance of the two-mode one-time-evaluation model is not as far as the other two models.

Simulation results of tolerable excess noise versus transmission distance is displayed in Fig.10, solid line, dashed line and pointed line are results under variance ${V}$ of 40, 20 and 4 respectively. Red lines denotes the three-mode one-time-calibration model, blue lines denotes two-mode one-time-calibration model and black lines denotes the original evaluation model. In the case of variance ${{V} = 20}$ and ${{V} = 4}$, the tolerable excess noise is almost the same for three-mode model and original two-time-evaluation model, both higher than the two-mode model. With variance ${{V} = 40}$, the original calibration model appears to have little higher tolerable excess noise than the other models.

\begin{figure}[t]
\centerline{\includegraphics[width=9.5cm]{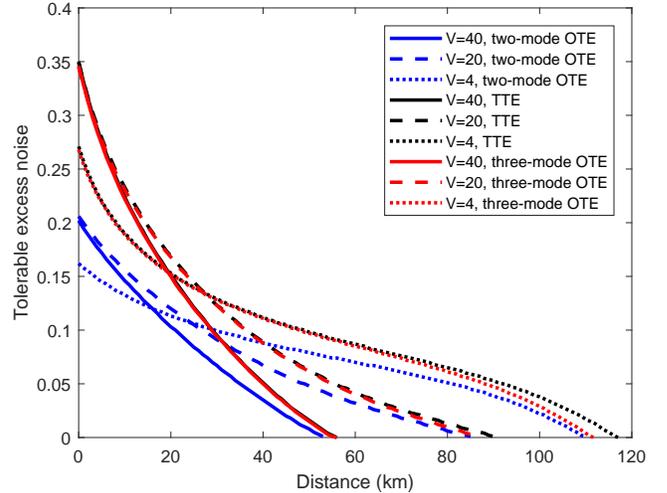}}
\caption{Tolerable excess noise as a function of transmission distance with the original two-time-evaluation model and the two one-time-evaluation models under different variances ${{V} = 40, 20, 4}$ in the finite-size regime. The electronic noise ${{v_{ele}} = 0.01}$, the limited detection efficiency ${{\eta _d} = 0.6}$ and the reconciliation efficiency ${\eta  = 0.956}$.}\label{only886644TEN}
\end{figure}

\begin{figure*}[t]
\centerline{\includegraphics[width=15cm]{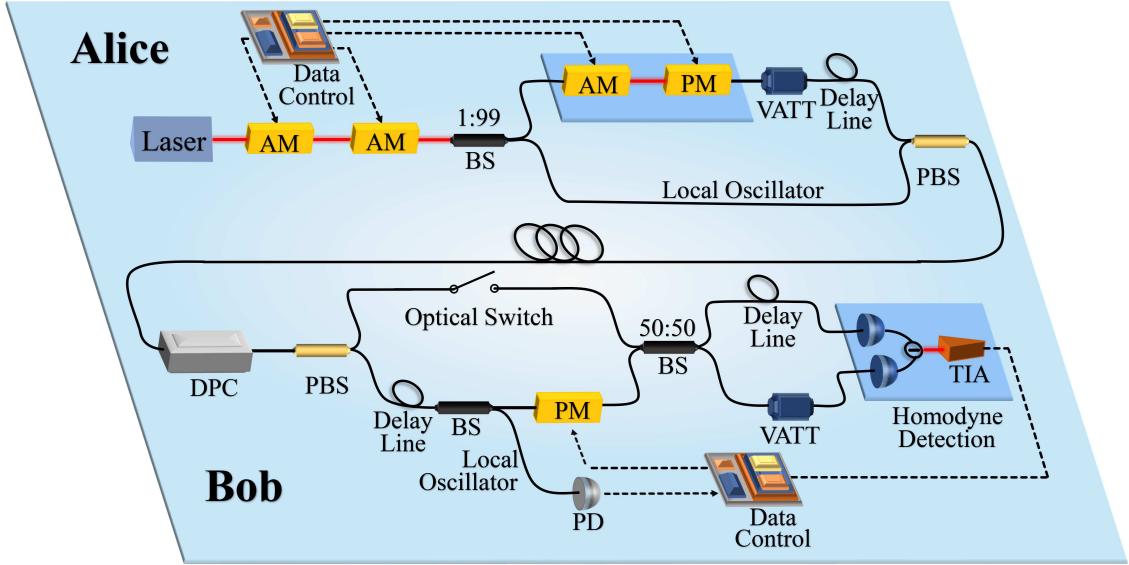}}
\caption{Optical layout of the experiment. Alice uses two high extinction ratio modulators and amplitude and phase modulator to prepare Gaussian distributed coherent states. Combined with strong local oscillator, the signals are multiplexed using polarizing beamsplitter and then sent through the quantum channel. The states are demultiplexed on Bob¡¯s side after the polarization compensation from the active dynamic polarization controller. The signal and local oscillator interfere on a shot-noise-limited balanced pulsed homodyne detector while the LO also attributes to the clock and data synchronization. Laser: continuous-wave laser; AM: amplitude modulator; PM: phase modulator; BS: beamsplitter; VATT: variable attenuator; PBS: polarizing beamsplitter; DPC: dynamic polarization controller; PD: photodetector.}\label{optical_layout_2}
\end{figure*}

\section{\label{sec:5} Experimental demonstration of one-time-calibration model}

We implement a proof-of-principle experiment to verify the feasibility of the one-time-calibration model. Since the three-mode model is outperforming than the two-mode model analysed in Sec.~\ref{sec:4}, three-mode one-time-evaluation model is used in the following. The schematic diagram of the complete optical layout is delineated in Fig. 11. In the experiment, a coherent-state protocol with a homodyne detection technique is adopted, where the legitimate party Alice generates coherent states from the laser and the other legitimate party, Bob, uses a homodyne detector to randomly measure one of the quadratures of the electromagnetic field. The experiment employs a 49.85 km fiber with a total channel loss of 11.62 dB.

In this PM model, the standard 1550nm telecom laser followed with two high-extinction amplitude modulators supplies 40ns coherent optical pulses which correspond to a duty circle of 20\% with a frequency of 5MHz. The pulses are then separated by the 1:99 beam splitter where the majority of them is treated as LO, the rest of the pulses are modulated by an amplitude modulator and an phase modulator subsequently so that a centered Gaussian distribution can be achieved. After the proper attenuation which can optimize the modulation variance and delay line, the signal is polarization and time multiplexed with the LO.
At Bob's side, the incoming signal of both the signal pulses and the LO pulses are first compensated by the dynamic polarization controller (DPC) to offset the polarization drifts. The polarization extinction ratio after the DPC maintains at a high level which attributes to separate the signal from the LO. Another delay line is used on Bob's side to compensate the time delay. The manipulations of the LO can be more complicated on Bob's side. 10\% of the LO is used for clock synchronization, data synchronization and LO monitor. The phase modulator in the LO path is responsible for phase compensation and random selection of the measured optical quadrature. An 80MHz shot-noise-limited balanced pulsed homodyne detector are then applied to detect the quantum signal.

One optical switch is adopted in the signal path in the receiver side, which corresponds to the corresponding EB model presented in Sec.~\ref{sec:3}. During the experiment, the SNU calibration process and the key distribution process are performed alternately. When the SNU calibration process is on, the optical switch is disconnected. When the optical switch is switched on, the key distribution process is executed.

Note that when the one-time calibration technique is used, the electronic noise no longer needs to be measured, and thus only one optical switch in the signal path of the receiver is enough to convert between SNU calibration and key distribution. In order to make a comparison, the experimental setup is changed a little, as another optical switch is placed in the LO path at the receiver end; in this way, the conventional calibration scheme can be used. In the process of SNU calibration, the optical switches in both the signal path and the LO path are first both switched off, and then the optical switch in the signal path is switched on. After the SNU calibration, the optical switch in the signal path is also switched on, and the key distribution process continues.

\begin{figure}[t]
\centerline{\includegraphics[width=9.5cm]{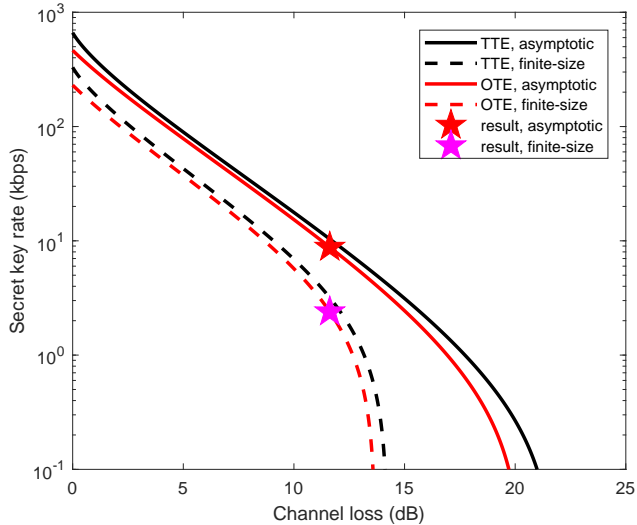}}
\caption{Secret key rate as a function of channel loss with the conventional two-time-calibration model and the one-time evaluation model in both the asymptotic regime and the finite-size regime. The red solid star is the mean secret key rate in our experiment under the asymptotic regime, while the purple star is the mean secret key rate in our experiment under the finite-size regime.}\label{SKR_exp_1}
\end{figure}

The calibration scheme used in this experiment offers several advantages over the conventional calibration procedure. The LO power is certainly retained, since the optical switch is no longer a necessity in the experimental implementation. This can be crucial when the transmission distance reaches a certain value, since the power of the LO is required to attain a certain level such that the quantum signal can be properly amplified during interference in the homodyne detection. The one-time calibration scheme also offers a simpler implementation and easier data control, as we need only to control one optical switch to complete the SNU calibration and the key distribution, and this is also suitable in the LLO scenario.

The basis sifting and parameter estimation in the postprocessing have low computational complexity, and thus can be implemented on a CPU. Multidimensional reconciliation and multiedge-type Low-Density Parity-Check codes are combined to perform information reconciliation, and this is suitable for achieving high efficiency at low SNRs~\cite{Leverrier_PRA_2008, Richardson_honor_2002}. The privacy amplification is implemented by using a hash Toeplitz function~\cite{Krawczyk_honor_1994, Fung_PRA_2010}. A high reconciliation efficiency inevitably results in a low frame error rate, and so this trade-off needs to be carefully mapped out. In this paper, the optimal reconciliation efficiency of 95.01\% obtained by using a rate-adaptive reconciliation protocol~\cite{Wang_QIC_2017, Zhou_Applied_2019} is investigated; this is conducive to maximizing the secret key rate.

The final secret key rates versus the channel loss in decibels obtained using the one-time evaluation method and the two-time evaluation method are depicted in Fig. 12 for both the asymptotic scenario and the finite-size scenario. The secret key rate of the one-time calibration model in the asymptotic regime is 11.62 kbps, and the secret key rate in the finite-size regime is 2.39 kbps. The secret key rate of the one-time calibration model is in fact no higher than that for the conventional model. This is due to the fact that the one-time evaluation model treats the electronic noise as a channel loss in the secret-key-rate calculation; however, with this negligible sacrifice, we manage to achieve a seamless switchover between the SNU calibration stage and the key distribution stage by using only one optical switch. This makes real-time monitoring of the SNU possible and will surely contribute to future commercial implementations. Besides, the SNU estimated by using the one-time evaluation model is more precise because of the reduced number of variables that cause statistical fluctuations, which provides a tighter bound on the secret key rates. Even with a slight diminution, the one-time calibration model still provides a comparable secret key rate.

\section{\label{sec:6} Conclusion}
In this paper, we propose a shot-noise unit calibration procedure that requires only one step to calibrate the shot-noise unit. We derive a complete entanglement-based model that corresponds to the proposed one-time calibration procedure and provide a full analysis of its performance. Complete experimental implementations based on a coherent-state homodyne scheme are conducted to experimentally verify its feasibility. We first review the conventional shot-noise unit calibration procedures and sketch the limitations of the existing model, and propose a one-time calibration method that can extend security to a more general noise environment where the relative-intensity noise can be addressed properly. This method will not only simplify the evaluation procedures but also make the estimated shot-noise unit more accurate, which makes it closer to an actual commercial implementation.

We perform a security analysis against arbitrary collective attacks, and secret-key-rate calculations for a two-mode version of the entanglement-based model, and then put forward a three-mode-calibration entanglement-based model by permuting the order of beam splitters, which further improves the behavior of the model. We show, further, the performance of the one-time calibration model in the finite-size regime. Our results from both the theoretical and the experimental perspective demonstrate that the one-time calibration model can essentially achieve nearly the same performance as the conventional calibration model, especially for the three-mode model in both the asymptotic-limit regime and the finite-size regime. Our proposal provides a better alternative method for the shot-noise-unit evaluation procedure.

It is worth noting that the proposed one-time calibration method could also have a significant impact in other continuous-variable quantum information fields using shot-noise unit calibration, such as quantum digital signatures~\cite{Croal_PRL_2016}, quantum secret sharing~\cite{Qi_PRA_2019}.

\begin{acknowledgments}
This work was supported in part by the Key Program of National Natural Science Foundation of China under Grants 61531003, the National Natural Science Foundation under Grants 61427813, China Postdoctoral Science Foundation under Grant 2018M630116, and the Fund of State Key Laboratory of Information Photonics and Optical Communications.
\end{acknowledgments}

%\bibliographystyle{unsrt}
%\bibliography{Reference333}

\end{document}